\documentclass[aps,pre, preprint,superscriptaddress,amsmath,amssymb,floatfix,showkeys]{revtex4-1}

\usepackage{graphicx}
\usepackage{times}
\usepackage{caption}
\usepackage{graphicx,amsmath,amssymb,amstext,amsfonts,color}
\usepackage{hyperref}
\usepackage{movie15}
\usepackage{color}
\usepackage[bottom]{footmisc}
\usepackage{natbib}

\usepackage{amsmath}
\usepackage{amssymb}
\def\D{\Delta}
\def\d{\delta}
\def\r{\rho}
\def\p{\pi}
\def\a{\alpha}
\def\g{\gamma}
\def\ra{\rightarrow}
\def\s{\sigma}
\def\b{\beta}
\def\e{\epsilon}
\def\G{\Gamma}
\def\om{\omega}
\def\l{\lambda}
\def\f{\phi}
\def\w{\psi}
\def\m{\mu}
\def\t{\tau}
\def\c{\chi}

\usepackage{times}
\DeclareMathOperator{\bx}{\mathbf{x}}

\begin{document}

\title{Evolutionary dynamics from deterministic microscopic ecological processes: A toy model for evolutionary processes}
\author{Vaibhav Madhok}
\email{vmadhok@gmail.com}
\affiliation{Department of Physics, Indian Institute of Technology Madras, Chennai, India 600036} 

\def\T{\Theta}
\def\D{\Delta}
\def\d{\delta}
\def\r{\rho}
\def\p{\pi}
\def\a{\alpha}
\def\g{\gamma}
\def\ra{\rightarrow}
\def\s{\sigma}
\def\b{\beta}
\def\e{\epsilon}
\def\G{\Gamma}
\def\om{\omega}
\def\l{\lambda}
\def\f{\phi}  
\def\w{\psi}
\def\m{\mu}
\def\t{\tau}
\def\c{\chi}
\begin{abstract}

The central goal of a dynamical theory of evolution is to abstract the mean evolutionary trajectory in the trait space by considering ecological processes at the level of the individual.
In this work, we develop such a theory for a new class of deterministic individual based models describing individual births and deaths, which captures the essential features of standard stochastic individual-based models and become identical with the latter under maximal competition. The key motivation is to derive the canonical equation of adaptive dynamics from this microscopic ecological model, which can be regarded as a ``toy model" for evolution, in a simple way and give it an intuitive geometric interpretation. Another goal is to study evolution and sympatric speciation under ``maximal" competition.
We show that these models, in the deterministic limit of adaptive dynamics, lead to the same equations that describe the unraveling of the mean evolutionary trajectory as those obtained from the standard stochastic models. We further study 
  conditions under which these models lead to evolutionary branching and find them to be similar with those obtained from the standard stochastic models.
 We find that though deterministic models result in a strong competition that leads to a speed up in the temporal dynamics of a population cloud in the phenotypic space as well as an increase in the rate of generation of biodiversity, it   
 does not seem to result in an absolute increase in biodiversity as far as total number of species are concerned. Hence, the ``toy model"
   essentially captures all the features of the standard stochastic model.
 Interestingly, the notion of a fitness function does not explicitly enter in our derivation of the canonical equation, thereby advocating a mechanistic view of evolution based on fundamental birth-death events where fitness is a derived quantity rather than a fundamental ingredient.
We illustrate our work with the help of several examples and qualitatively compare the rate of unraveling of evolutionary trajectory and generation of biodiversity for the deterministic and standard individual based models by showing the motion of population clouds in the trait space.

\end{abstract}


\maketitle
\vskip 0.5 cm
\section{Introduction}
\label{intro}
 Evolution is essentially a birth-death process involving inheritance, mutations and selection. Studying how populations evolve in long term involves describing the behavior of individual constituents of the population, the stochastic birth death processes, in order to address questions about origin of diversity, speciation
 and predictability of evolution \cite{doebeli2011, 10.7554/eLife.23804}. 
  In order to have such a view, we need to take into account certain underlying ecological mechanisms governing birth and death processes, like the availability of food, fertility of individuals and the interaction among organisms \cite{doebeli2011, dieckmann_doebeli1999, dieckmann_doebeli2004, geritz_etal1998}. 
  
  A dynamic theory of evolution aims to describe the long-term phenotypic distribution in terms of underlying microscopic ecological processes  \cite{dieckmann_law1996} where
  the birth and death rates of the individual depend on its phenotypic composition as well as its interaction with the ``environment". The ``environment" consists of both the ``abiotic" environment like temperature, climate etc. and the ``biotic" 
 environment which can be, for example, the types of all other individuals with which it interacts and the competition it faces with regards to resources such as food and shelter. Another example of the ``biotic" environment is the presence of ``prey" or ``predator" type of individuals in the population \cite{doebeli2011}.

 Adaptive dynamics is one such framework that incorporates frequency dependent selection originating from microscopic ecological interactions and competition among species \cite{doebeli2011}.  Under this framework, evolution is regarded as a continuous trajectory in the phenotypic space that unravels as rare mutants with a higher ``fitness" invade a resident population.
 The evolutionary dynamics in the above framework is described by
 the ``canonical equation" \cite{dieckmann_law1996}, which, for one dimension, is given by
 \begin{align}
 \label{AD_basic}
 \frac{d{\textbf{x}}}{dt} = \gamma \frac{\partial f(\bold{x}, \bold{y})}{\partial \textbf{y}}_{\big | \bold{y}=\bold{x}}.
\end{align}  
     
 Here $\gamma$ is a constant of proportionality that scales the rate of evolution. The function, $ f(\bold{x}, \bold{y})$, is the measure of fitness of a rare mutant, $y$, in a monomorphic population with a resident trait $x$ \cite{dieckmann_law1996}. 
The derivative of the invasion fitness function, $f(\bold{x}, \bold{y})$, with respect to the mutant phenotype coordinate, $y$, gives the direction of motion of the evolutionary trajectory. Mutations cause the occurrence of rare types which can possibly invade the resident depending on their relative growth rate with respect to the resident which is given by the selection gradient (RHS of Eq \ref{AD_basic}).    

 The derivation of the canonical equation takes into account the stochastic nature of the birth-death process
and is based on the ecological processes influencing the birth and death at the level of an individual \cite{dieckmann_law1996, champagnat_etal2008,champagnat_etal2006}. Such a formulation assumes that mutations are infinitesimal, they occur very rarely and the resident population is taken to be of infinite size \cite{champagnat_etal2008}. However, numerical evidence suggests that adaptive dynamics agrees with the individual based simulations 
despite of the fact that the latter involves finite sized mutations and populations. 
The canonical equation 
which describes the dynamics of a monomorphic population in the trait space driven by mutations, selection and invasion follows the centre of mass of the population cloud in the individual based simulations. The key focus of our study is to better understand this correspondence and offering a simpler derivation of the canonical equation
from individual based models.

For this purpose, we propose a new class of individual based models that, in the deterministic limit, preserve the canonical equation of adaptive dynamics and give qualitatively similar conditions for evolutionary branching as those obtained with the standard individual based models. 
We derive the canonical equation from this individual based model and give it a geometrical interpretation. As we shall see, the derivation does not explicitly involve the fitness function. We show how to think of evolution without a fitness function as a birth-death process where fitness is a derived quantity rather than a base ingredient \footnote{The author acknowledges Michael Doebeli for this perspective}. 
Therefore, our model captures all the essential features of standard stochastic individual based models and therefore may be regarded as a toy theory for evolutionary studies that gives us an expression for the mean evolutionary trajectory in a more intuitive and simpler fashion. Alternatively, the models we propose give us a description of evolution under maximal competition and therefore can be useful to study evolution under certain abiotic conditions like, for example, the high competition limit of standard stochastic models. The ecological significance of these models is discussed. For example, with the help of numerical simulations \footnote{See Supplemental Material at [URL will be inserted by publisher] for examples and movies of simulations}, we demonstrate how these models lead to faster motion of the population cloud in the trait space. As another example of our model, we discuss  the issue of the predictability of evolution by considering individual based models in high dimensional phenotypic spaces and demonstrate the ``evolutionary butterfly effect". 

Throughout the paper, we use the acronym ``DMEP" to represent Deterministic Microscopic Ecological Process, the new class of individual based models that we introduce. We will use ``SSIBM" to represent ``Standard Stochastic Individual Based Models", the usual way to implement microscopic birth-death processes using the Gillespie algorithm \cite{gillespie1976}.

\section{From microscopic individual based models to macroscopic deterministic adaptive dynamics}
We begin with the widely studied logistic model \cite{may1976}, 
 
 \begin{align}
\label{logistic}
 \frac{\partial N(\bold{x}, t)}{\partial t} = r N(\bold{x}, t)\bigg( 1 - \frac{\int \alpha(\bold{x}, \bold{y}) N (\bold{y}, t) dy}{K(\bold{x})}\bigg).
\end{align}
We will call this the ``canonical" model, as later on we will also be interested in a general family of such models to describe aspects of ecology like the population dynamics.  
Here $N(\bold{x}, t)$ is the population density of the phenotype $\bold{x}$ at time $t$ and $K(\bold{x})$ is the carrying capacity of the population consisting entirely of $\bold{x}$ individuals. 
The competition between individuals of phenotypes $\bold{x}$ and $\bold{y}$ is given by the competition kernel $\alpha(\bold{x}, \bold{y})$ and an individual of phenotype $\bold{x}$ faces competition with an effective density $\int \alpha(\bold{x}, \bold{y}) N (\bold{y}, t) dy$. We set the intrinsic growth parameter $r$ and $\alpha(\bold{x},\bold{x})$ to be equal to $1$. We consider $\bold{x}$ and $\bold{y}$ to be vectors of dimension $d$ describing a multidimensional phenotypic space.
This will be our primary running example throughout the paper. Our goal is to extract information from the above dynamics about how phenotypic traits in a population evolve over time, or the evolution in the trait space. Adaptive dynamics is a framework that gives us this information from the ecology and population dynamics described above.
  
The canonical equations of adaptive dynamics, describing the evolution in the phenotypic space,  can be derived by considering the stochastic individual based model corresponding to \ref{logistic} in the limit of rare mutations, small mutational effects and infinite population sizes.   
Under these assumptions, Dieckmann and Law showed that adaptive dynamics is the first order approximation of the mean path averaged over infinitely many realizations of the stochastic simulations obtained from the individual based model. In order to derive the adaptive dynamics of trait $\bold{x}$, we assume a monomorphic (Dirac-delta distribution) resident population in trait $\bold{x}$  with a globally stable equilibrium density given by $K(\bold{x})$ independent of the dimension $d$ of $\bold{x}$. A key ingredient to obtain the ODEs describing the adaptive dynamics is the \textit{invasion fitness} function whose gradient determines the direction of selection forces and hence of the deterministic evolutionary trajectory.
The \textit{invasion fitness} function is the difference between the per capita birth and per capita death rates of a rare mutant in a monomorphic population. 
The \textit{invasion fitness}, for the logistic map, of a rare mutant $\bold{y}$ is its per capital rate of growth in a resident population with phenotype $\bold{x}$ and is
given by
 \begin{align}
 \label{fitness}
f(\bold{x}, \bold{y}) = 1 - \frac{ \alpha(\bold{x}, \bold{y}) K(\bold{x})}{K(\bold{y})}.
\end{align}
In the above example for the logistic equation, the selection due to frequency dependence influences the death rates while the birth rates are neutral. In general, birth rates might also depend on interactions between individuals and the fitness function will have a more general form involving competition kernels for both birth and death.

From the \textit{invasion fitness}, we can derive the selection gradient to be
 \begin{align}
\label{sg1} 
s_i(\bold{x}) = \frac{\partial f(\bold{x}, \bold{y})}{\partial y_i}_{\big | \bold{y}=\bold{x}} =  - \frac{\partial \alpha(\bold{x}, \bold{y})}{\partial y_i}_{\big | \bold{y}=\bold{x}} + \frac{\partial K(\bold{x})}{\partial x_i}\frac{1}{K(\bold{x})}.
\end{align}
Finally, the canonical equation of adaptive dynamics in terms of the selection gradient is given by
 \begin{align}
 \label{AD1}
 \frac{d\bold{x}}{dt} = M(\bold{x}). s(\bold{x}).
\end{align}
Equation (\ref{AD1}) is a system of $d$ dimensional ODEs that describe a trajectory that may converge to a fixed point, a limit cycle, a quasi periodic orbit or exhibit chaos depending on the nature of the selection gradient $s(x)$ and the initial conditions.
 Here $M(x)$ describes the mutational process and for simplicity we assume it to be a $d \times d$ identity matrix. 

 Individual based models describe dynamics of population at a microscopic level and give a stochastic description of interactions between discrete individuals that possess multi-dimensional adaptive traits.
  Such a microscopic description is more general and in a sense more fundamental as compared to its deterministic macroscopic approximation like the adaptive dynamics which can be derived from it under suitable assumptions. 
   Adaptive dynamics, as given by Eq. (\ref{AD1}), is a first order derivative of the invasion fitness function with respect to the mutant phenotype, $\bold{y}$, evaluated at $\bold{y} = \bold{x}$. As was shown in \cite{ISPOLATOV201697}, there are infinitely many invasion fitness functions with different competition kernels that give the same adaptive dynamics. Thus, for a given deterministic adaptive dynamics, the choice of the competition kernel $\alpha(\bold{x}, \bold{y})$ in the corresponding individual based model is not unique \cite{ISPOLATOV201697}. 

 In order to simulate individual based models, we adapt the Gillespie algorithm \cite{gillespie1976} for simulating chemical reactions to population biological stochastic processes.  Individuals are treated as particles, and birth and death events as chemical reactions. In the deterministic logistic model, the birth rate of phenotype $i$, $\rho_{i}$, with trait value $\bold{x_i}$ is 1. When a birth event happens, a new individual is added to the population with a phenotype that is offset from the 
 parent by a small mutation, which happens with probability $\mu$, chosen from a uniform distribution with a small amplitude. 
 Throughout this paper we choose $\mu=1.0$.  
 Its death rate, ${\delta_{i}}$ is given by $N_{eff}(i)/K(\bold{x_i})$, where $N_{eff}$ is the effective density experienced by phenotype $\bold{x}$. For the individual based model, if there are $N$ individuals, $\bold{x}_1$, ..., $\bold{x}_N$, at time $t$ then the effective density experienced by individual $i$ is
   \begin{align}
 \label{Neff} 
 N_{eff}(i) = \sum_{j \neq i} \alpha({\bold{x}_j, \bold{x}_i}),
\end{align}
and, therefore, the death rate is given by
\begin{align}
 \label{Neff}
 d_{eff}(i) =  N_{eff} (i) /K(\bold{x}_i).
\end{align}

The individual based simulations are then implemented stochastically in the familiar way by performing one birth or death event \cite{gillespie1976} that constitutes one computational step that advances the system from time $t$ to $t + \Delta t$. In the above example for the logistic equation, the selection due to frequency dependence influences the death rates while the birth rates are neutral. In general, both the birth as well as the death rates can be affected by frequency dependent interactions among individuals described by competition terms like $\alpha_{birth}({\bold{x}_j, \bold{x}_i)}$ and $\alpha_{death}({\bold{x}_j, \bold{x}_i)}$. The algorithm consists of assigning each individual $i$  a 
constant reproduction rate $\rho_{i}=1$ and a death rate $\delta_{i}=\sum_{\j\neq \i}
\alpha(\bx_{i},\bx{_j})/K(\bx_{i})$, as defined by logistic ecological dynamics.
 The total rate is
given by the sum of all individual rates $U=\sum_{i}
(\rho_{i}+\delta_{i})$. A particular ``event", birth or death, is chosen 
at random with probability equal to the rate of this event divided by the total probability
rate $U$. If a birth event is chosen, a new individual is added to the population with a phenotype that is offset from the parent by a small mutation chosen from a uniform distribution with a small amplitude. In case of a death event, the individual is removed from the population.

\section{Deterministic microscopic ecological processes}

In this section we introduce a new class of ``deterministic" microscopic ecological processes, DMEP,
which consists of choosing the individual with the maximum birth/death rate
and adding/eliminating it as the case may be. 
 For example, if both the birth as well as the death rates are affected by frequency dependent interactions among individuals described by competition terms like $\alpha_{birth}({\bold{x}_j, \bold{x}_i)}$ and $\alpha_{death}({\bold{x}_j, \bold{x}_i)}$, when the birth event is chosen, we select the individual with the maximum birth rate, $\rho_{i}$, and a new individual is added to the population with a phenotype that is offset from the parent by a small mutation chosen from a uniform distribution with a small amplitude. In case of the death event, we select $max(\delta_1, \delta_2, ..., \delta_n)$ and eliminate that individual. The case when birth/death rates are neutral can be treated as special cases and any individual can be added/removed at random. As we shall see, this simple modification to the standard stochastic models yields gives a very simple derivation of adaptive dynamics and helps us interpret it geometrically. 
 
 We now analyse this particular stochastic model and its relation to adaptive dynamics. First we will start from the ecological model and discuss how the SSIBM goes to DMEP in the limit of strong competition and its consequences for the resulting adaptive dynamics . In section \ref{deriv}, we start with DMEP and derive the canonical equation of adaptive dynamics from first principles and give it a geometrical interpretation.

\subsection{DMEP: From ecology to adaptive dynamics under strong competition}

In order to understand DMEP and their dynamics, we first consider a logistic model of the form

\begin{align}
\label{modilo}
 \frac{\partial N(\bold{x}, t)}{\partial t} =  N(\bold{x}, t)\Bigg( 1 - \bigg(\frac{\int \alpha(\bold{x}, \bold{y}) N (\bold{y}, t) dy}{K(\bold{x})}\bigg)^m\Bigg).
\end{align}
The above model reduces to Eq. \ref{logistic} for $m=1$.
We assume the resident to be at the equilibrium population density, given by the Dirac delta function,
$N(x') = K(x)\delta(x-x')$. Then the per capita growth of a mutant $y$ is given by

\begin{align}
\label{pcmod}
\frac{1}{N(\bold{y}, t)} \frac{\partial N(\bold{y}, t)}{ \partial t} =\Bigg( 1 - \bigg(\frac{\int \alpha(\bold{x'}, \bold{y}) K(x) \delta(x-x') dx'}{K(\bold{y})}\bigg)^m\Bigg).
\end{align}
Therefore, the invasion fitness, as defined above is given by
\begin{align}
\label{IFmod}
f(\bold{x}, \bold{y}) =\Bigg( 1 - \bigg(\frac{ \alpha(\bold{x}, \bold{y}) K(x) }{K(\bold{y})}\bigg)^m\Bigg).
\end{align}
From this, we can derive the selection gradient to be
\begin{widetext}
 \begin{align}
\label{msg1}
s_i(\bold{x}) = \frac{\partial f(\bold{x}, \bold{y})}{\partial y_i}_{\big | \bold{y}=\bold{x}} =  m \bigg(\frac{ \alpha(\bold{x}, \bold{y}) K(x) }{K(\bold{y})}\bigg)^{m-1}_{\big | \bold{y}=\bold{x}} \bigg(-\frac{\partial \alpha(\bold{x}, \bold{y})}{\partial y_i}_{\big | \bold{y}=\bold{x}} + \frac{\partial K(\bold{x})}{\partial x_i}\frac{1}{K(\bold{x})}\bigg), 
\end{align}
\end{widetext}
which simplifies to
\begin{widetext}
\label{msg}
\begin{align}
s_i(\bold{x}) = \frac{\partial f(\bold{x}, \bold{y})}{\partial y_i}_{\big |\bold{y}=\bold{x}} = m\bigg(- \frac{\partial \alpha(\bold{x}, \bold{y})}{\partial y_i}_{\big | \bold{y}=\bold{x}}+\frac{\partial K(\bold{x})}{\partial x_i}\frac{1}{K(\bold{x})}\bigg).
\end{align}
\end{widetext}
Therefore, the above selection gradient is $m$ times the selection gradient given in Eq. \ref{sg1}.

Finally, the canonical equation of adaptive dynamics in terms of the selection gradient is given by
 \begin{align}
 \label{AD2}
 \frac{d\bold{x}}{dt} = m M(\bold{x}). s(\bold{x}),
\end{align}
which gives the same set of ODEs up to the scaling constant $m$ as obtained previously with the canonical logistic model. We also note that the Eq.\ref{AD2} evolves $m$ times faster than Eq. \ref{AD1}.
Though derivation of Eq. \ref{AD2} works for unimodal populations, our simulations in the next section suggest that a similar correspondence results even after evolutionary branching leads to multi-modal populations. In fact, Eq. \ref{modilo} represents a family of logistic like models that give the same adaptive dynamics upto a scaling factor. In the next section, we shall see evolutionary branching for such models is also very similar to each other. To derive the canonical equation and interpret it, we will be more interested in the deterministic limit (large $m$) of the this equation as shown below.

In constructing the individual based models for the logistic model in Eq. \ref{modilo}, 
we note that the birth rates remain constant  at $\rho_{i}=1$, while the  death rate for the individual $i$ is given by $\delta_{i}=\big(\sum_{\j\neq \i}
\alpha(\bx_{i},\bx{_j})/K(\bx_{i})\big)^m$. Therefore, the array of these death rates can be expressed as  $(\delta_1^m, \delta_2^m, ..., \delta_n^m)$
where $(\delta_1, \delta_2, ..., \delta_n)$ are the death rates of individuals corresponding to the ``canonical" map. Let $k$ be the individual with the highest death rate. Thus $max(\delta_1^m, \delta_2^m, ..., \delta_n^m) = \delta_k^m$.
When $m$ is sufficiently large, we have $\delta_k^m >>\delta_i^m$ for all $i \neq k$. Therefore the probability of choosing $k$ to die is close to 1. But this is exactly what we do in implementing DMEP.
We deterministically pick the individual with the highest death rate and eliminate it. Therefore, the DMEP  can be considered to be an individual based simulation of the standard birth death process governed by the logistic model given by Eq. \ref{modilo}. The value of $m$ in Eq. \ref{modilo} should be sufficiently large 
such that, we satisfy, $\delta_k^m >>\delta_i^m$ for all $i \neq k$ so that the process of selecting an individual to die becomes deterministic.
We have already shown that the individual based simulations based on this model, in the deterministic limit, give rise to exactly the same equations of adaptive dynamics as the 
``canonical" model up to the scaling constant $m$.   

\section{Conditions for diversification in one dimension}

Adaptive diversification has been extensively studied at singular points of the dynamics that also happen to be evolutionarily unstable \cite{doebeli2011, geritz_etal1998}, and 
\cite{metz_etal1992}. The canonical equation of adaptive dynamics has the form
\begin{align}
 \label{CAD}
 \frac{dx_i}{dt} = g_i(x),
\end{align}
    where the functions $g_i$ are given as
    \begin{align}
 \label{AD}
 \begin{pmatrix} g_i(x) \\ ... \\ g_d(x)\end{pmatrix} = \begin{pmatrix}  \frac{\partial f(x, y)}{\partial y_i}_{\big | y=x}  \\ ... \\  \frac{\partial f(x, y)}{\partial y_d}_{\big | y=x} \end{pmatrix}.
\end{align}
Singular points of the dynamics are the points, $\bold{x}^{*}$ at which the R.H.S. of Eq. (\ref{CAD}) is zero. 
Singular point, $\bold{x}^{*}$ is called \textit{convergence stable} if the dynamics starting at all points sufficiently close to it converge to $\bold{x}^{*}$ eventually. This occurs when the Jacobian at the singular point, $J(\bold{x}^*) <0$.
These points are of great importance in adaptive dynamics as they can be potentially the points at which diversification occurs. 

For scalar traits, this can be seen by Taylor expanding the invasion fitness function with respect to the mutant $y$ to the second order, we get
\begin{widetext}
\begin{align}
 \label{TEF}
 f(x,y) = f(x, x) + \frac {\partial f(x, y)}{\partial{y}}_{\big |y=x}.(y-x)+  \frac{\partial^2 f(x, y)}{\partial y^2}_{\big | y=x}.\frac{(y-x)^2}{2}.
\end{align}
\end{widetext}
The first term on the RHS, $f(x, x)$ is zero for all $x$.
Usually, the evolutionary dynamics can be accurately described by the first order term in (\ref{TEF}) or, in other words, by the canonical equation of adaptive dynamics and we do not need to consider higher orders terms.
It is only in the neighbourhood of singular points, we have $ \frac {\partial f(x, y)}{\partial{y}}_{\big | y=x} \rightarrow 0$, the second order term in Eq. (\ref{TEF}) becomes significant. In particular, at the singular point we have $ \frac {\partial f(x, y)}{\partial{y}}_{\big | y=x} = 0$ and if $ \frac{\partial^2 f(x, y)}{\partial y^2}_{\big | y=x} < 0$, no nearby mutants can invade the resident population that is monomorphic and we have the conditions for evolutionary stability. In contrast,  $ \frac{\partial^2 f(x, y)}{\partial y^2}_{\big | y=x} > 0$ is the condition for evolutionary instability, or potential evolutionary branching points, as the mutant now can potentially invade the resident.

In order to study diversification under DMEP, we need to explore higher orders of the derivative of the  fitness function with respect to
the mutant trait.
It is easy to see that the fitness functions derived from Eq. \ref{logistic} and Eq. \ref{modilo} yield the same conditions for diversification. For the Eq. \ref{logistic} we have

\begin{align}
\label{div_1}
 \frac{\partial^2 f(x, y)}{\partial y^2}_{\big | y=x}   = -\frac{\partial}{\partial y}\Bigg(\frac{\partial}{\partial y}\bigg(\frac{\alpha(x, y) k(x)}{k(y)}\bigg)\Bigg)_{\big | y=x},
\end{align}

and for the invasion fitness function in Eq. \ref{IFmod}, the second derivative is
\begin{widetext}
\begin{equation}
\begin{aligned}
 \frac{\partial^2 f(x, y)}{\partial y^2}_{\big | y=x}   =-m(m-1)\bigg(\frac{\alpha(x, y) k(x)}{k(y)}\bigg)^{m-2}\bigg(\frac{\partial}{\partial y}\bigg(\frac{\alpha(x, y) k(x)}{k(y)}\bigg)\bigg)^{2}_{\big | y=x} \\ 
   - m\Bigg(\frac{\alpha(x, y) k(x)}{k(y)}\Bigg)\frac{\partial}{\partial y}\Bigg(\frac{\partial}{\partial y}\bigg(\frac{\alpha(x, y) k(x)}{k(y)}\bigg)\Bigg)_{\big | y=x}.
\end{aligned}
\end{equation}
\end{widetext}
At equilibrium, when we have $ \frac {\partial f(x, y)}{\partial{y}}_{\big | y=x} = 0$, the first term on the RHS is zero. And the second term coincides with Eq. \ref{div_1}. Therefore, the conditions for potential diversification (till the first branching point) are identical for both maps. Therefore, the family of logistic like models, parametrised by $m >1$, given by Eq. \ref{modilo} have the same conditions for diversification in one dimension. Since DMEP represents the large $m$ limit, it follows that conditions for diversification under DMEP is similar to that under SSIBM.

\section{Conditions for diversification in multiple dimensions} 

The above arguments are extended to higher dimensional systems and evolutionary branching is studied 
at convergence stable singular point of the multi dimensional system. In fact, convergence stability is often regarded as a necessary condition for evolutionary branching. In \cite{ISPOLATOV201697, ito_dieckmann2014}, it was shown that this is not the case. Higher dimensional version of Eq. (\ref{TEF}) is given by:
\begin{widetext}
\begin{align}
 \label{HTEF}
 f(\bold{x},\bold{y}) = f(\bold{x}, \bold{x}) +\Big[\frac {\partial f(\bold{x}, \bold{y})}{\partial{y_i}}, ..., \frac {\partial f(\bold{x}, \bold{y})}{\partial{y_d}} \Big]_{\big | y=x}  (\bold{y}-\bold{x}) +\frac{1}{2} (\bold{y}-\bold{x})^T \bold{H\bold(x)} (\bold{y}-\bold{x}),
\end{align}
\end{widetext}

where $H(\bold{x})$ is the Hessian matrix of second derivatives of the invasion fitness function with respect to the mutant trait values, $y$, evaluated at the resident trait value $x$:
\begin{align}
 H(\bold{x}) = \begin{pmatrix} \frac{\partial^2}{\partial y_1 \partial y_1} f(\bold{x}, \bold{y})_{\big | \bold{y}=\bold{x}} & ... & \frac{\partial^2}{\partial y_1 \partial y_d} f(\bold{x}, \bold{y})_{\big | \bold{y}=\bold{x}} \\...&...&...\\ \frac{\partial^2}{\partial y_d \partial y_1} f(\bold{x}, \bold{y})_{\big | \bold{y}=\bold{x}}&...&\frac{\partial^2}{\partial y_d \partial y_d} f(\bold{x}, \bold{y})_{\big | \bold{y}=\bold{x}}\end{pmatrix}_.
\end{align}
  
In this scenario, the first order term can become zero or be arbitrarily close to zero when the vector $(\bold{y}-\bold{x})$ is orthogonal or near orthogonal to the gradient, $\Big[\frac {\partial f(\bold{x}, \bold{y})}{\partial{y_i}}, ..., \frac {\partial f(\bold{x}, \bold{y})}{\partial{y_d}} \Big]_{\big | y=x}$. In other words,  the second order terms for trait values, $\bold{y}$, that lie orthogonal to the direction of the gradient of $f(\bold{x}, \bold{y})$ become significant regardless of the trajectory being in the vicinity of a singular point.  In particular,  if $ H(\bold{x})$ is negative definite, no nearby mutants can invade the resident population that is monomorphic and we have the conditions for evolutionary stability. When the Hessian matrix in the $d-1$ dimensional subspace orthogonal to the first order gradient is not negative we get the condition for evolutionary instability. The mutants residing in this $d-1$ plane can potentially invade the resident. Geometrically, the Hessian matrix associated with the $d-1$ subspace determines the local curvature of the $d-1$ dimensional plane orthogonal to the first order gradient. Depending on this curvature (the non negativity of the Hessian matrix), there may exist directions along which the invasion fitness function is a minima with respect to the mutant trait values and hence become evolutionary unstable.

  Adaptive dynamics, as given by Eq. (\ref{sg1}) and Eq. \ref{AD1}, is a first order derivative of the invasion fitness function with respect to the mutant phenotype, $\bold{y}$, evaluated at $\bold{y} = \bold{x}$. There are infinitely many  the invasion fitness functions with different competition kernels that give the same adaptive dynamics. Thus, for a given deterministic adaptive dynamics, the choice of the competition kernel $\alpha(\bold{x}, \bold{y})$ in the corresponding individual based model is not unique. 
To see this, let's reverse this procedure and derive an individual-based model from the system of equation corresponding to the given adaptive dynamics.
For example, using Eq. \ref{sg1}, the adaptive dynamics equations of Eq. \ref{AD1} can be expressed as
\begin{equation}
\label{Gen_AD}
\frac{dx_i}{dt} =  w_i(x) - u_i(x),
\end{equation}  
  with the first term coming from the competition kernel and the second term from the carrying capacity.
  Such division into $w$ and $v$ terms may look artificial from the mathematics view point (it's possible to include everything into $w_i$) but it follows from the underlying evolution model. We will consider examples (see supplementary material)
  of individual based models described by different competition kernels that give the same adaptive dynamics \cite{ISPOLATOV201697} as described by Eq. \ref{Gen_AD}.\\
\subsection{The ``canonical" individual based model}
In this section, we show that there exists a ``canonical" choice for the competition kernel, for which the associated Hessian matrix described above is negative definite and hence the individual-based model is expected to remain monomorphic. We start by expressing the canonical competition kernel, $\alpha(\bold{x}, \bold{y})$, as
\begin{align}
 \label{ACCC}
\alpha({\bold{x}, \bold{y}}) = \exp\Big(\sum_{i=1}^{d} w_i(\bold{x})(x_i - y_i)\Big),
\end{align}
  \begin{align}
 \label{P1CCC}
\frac{\partial \alpha({\bold{x}, \bold{y}})}{\partial y_i} = -w_i(\bold{x}) \alpha({\bold{x}, \bold{y}}),
\end{align}   
  and
\begin{align}
 \label{P2CCC}
\frac{\partial^2 \alpha({\bold{x}, \bold{y}})}{\partial y_i \partial y_j} = w_i(\bold{x}) w_j(\bold{x}) \alpha({\bold{x}, \bold{y}}),
\end{align}   
Using the above results, we find that the matrix entries of the Hessian of the invasion fitness function are
\begin{widetext}
\begin{align}
 \label{FH}
\frac{\partial^2 f({\bold{x}, \bold{y}})}{\partial y_i \partial y_j} = -\frac{\partial^2 \alpha({\bold{x}, \bold{y}})}{\partial y_i \partial y_j} \frac{K(\bold{x})}{K(\bold{y})} + \frac{\partial \alpha({\bold{x}, \bold{y}})}{\partial y_i} \frac{K(\bold{x})}{K(\bold{y})^2}\frac{\partial K(\bold{y})}{\partial y_j} +  \frac{\partial \alpha({\bold{x}, \bold{y}})}{\partial y_j} \frac{K(\bold{x})}{K(\bold{y})^2}\frac{\partial K(\bold{y})}{\partial y_i} \\+\alpha(\bold{x}, \bold{y}) \Big(\frac{K(\bold{x})}{K(\bold{y})^2} \frac{\partial^2 K({\bold{y}})}{\partial y_i \partial y_j} - 2\frac{K(\bold{x})}{K(\bold{y})^3}\frac{\partial K(\bold{y})}{y_i}\frac{\partial K(\bold{y})}{\partial y_j}\Big) \nonumber
\end{align}
\end{widetext}
Moreover, given the competition kernel, we calculate the elements of the Hessian of the invasion fitness function as
\begin{align}
 \label{FHE}
\mathbf H(\mathbf{x})_{ij} = &- w_i(\mathbf{x}) w_j(\mathbf{x}) + w_i(\mathbf{x}) u_j(\mathbf{x}) +w_j(\mathbf{x}) u_i(\mathbf{x}) - 2u_i(\mathbf{x})u_j(\mathbf{x})  +\frac{1}{K(\mathbf{x})} \frac{\partial^2 K({\mathbf{x}})}{\partial x_i \partial x_j}
\end{align}

This matrix and can be written as 
\begin{align}
 \label{FHEM1}
\mathbf{H}(\mathbf{x}) = &- \begin{pmatrix}  w_1 - u_1\\...  \\w_d - u_d\end{pmatrix} \begin{pmatrix}  w_1 - u_1, & ...,& w_d - u_d\end{pmatrix} - \begin{pmatrix}   u_1\\...  \\u_d\end{pmatrix} \begin{pmatrix}  u_1, & ...,&  u_d\end{pmatrix}+ \frac{\mathbf{H_K}(\mathbf{x})}{K(\mathbf{x})},
\end{align}
where $\mathbf{H_K}(\mathbf{x})$ is the Hessian of the carrying capacity, which is by assumption negative definite (and where we have omitted the argument $\mathbf{x}$ from the terms $w_i(\mathbf{x})$ and $u_i(\mathbf{x})$).

The first term in (\ref{FHEM1}) is a rank 1 matrix 
and its only non-zero eigenvalue $\l$ is negative:
\begin{align}
 \label{lambda}
\l = -\sum_{i=1}^d [w_i(\mathbf{x}) - u_i(\mathbf{x})]^2.
\end{align}

Similarly, the second term in (\ref{FHEM1}) is also a rank 1 matrix 
and its only non-zero eigenvalue $\l$ is negative:
\begin{align}
 \label{lambda1}
\l = -\sum_{i=1}^d [u_i(\mathbf{x})]^2.
\end{align}

As consequence, with the minimal competition kernel the Hessian of the invasion fitness function minimal competition kernel is always negative definite, independent of the current resident phenotype, and hence diversification is not expected anywhere along the trajectory of the adaptive dynamics.

 For an explicit example of carrying capacity function that we will also use in simulations, we will assume it to be of the form $K(x) = \exp (-\sum_i x_i^4/4)$. In general, our arguments will hold for any function of the form $K(x) = \exp (-\sum_i K_i |x_i|^n)$, which is unimodal and acts as a stabilizing component of selection for trait value $\bold{x} = 0$.
 
Evaluating this at the resident coordinate, $y=x$, with $\alpha(\bold{x},\bold{x}) = 1$ and using Eq. \ref{P1CCC} and \ref{P2CCC}, we get
\begin{widetext}
\begin{align}
 \label{FHE}
\frac{\partial^2 f({\bold{x}, \bold{y}})}{\partial y_i \partial y_j} = - w_i(\bold{x}) w_j(\bold{x}) + w_i(\bold{x}) x_j^3 +w_j(\bold{x}) x_i^3 - 2 x_i^3x_j^3 -3x_i^2\delta_{ij}
\end{align}
\end{widetext}
The above matrix and can be expressed as 
\begin{widetext}
\begin{align}
 \label{FHEM}
H(\bold{x}) = - \begin{pmatrix}  w_1(\bold{x}) - x_1^3 \\...  \\w_d(\bold{x}) - x_d^3\end{pmatrix} \begin{pmatrix}  w_1(\bold{x}) - x_1^3 & ...& w_d(\bold{x}) - x_d^3\end{pmatrix} - \begin{pmatrix}  x_1^3 \\...  \\ x_d^3\end{pmatrix} \begin{pmatrix}  x_1^3 & ...&  x_d^3\end{pmatrix}  +     \begin{pmatrix} -3x_1^2&...&0\\...&...&...\\0&...&-3x_d^2\end{pmatrix}.
\end{align}
\end{widetext}
The first term in the above is a rank 1 matrix 
and its only non-zero eigenvalue, $\l$, is given by:
\begin{align}
 \label{lambda}
\l = -\sum_{i=1}^d (w_i(\bold{x}) - x_i^3)^2,
\end{align}
which is negative for all values of $\bold{x}$. 
Similarly, the second term in the above is a rank 1 matrix 
and its only non-zero eigenvalue, $\l$, is given by:
\begin{align}
 \label{lambda}
\l = -\sum_{i=1}^d [x_i^3]^2,
\end{align}
which is negative for all values of $\bold{x}$.

The third term is a negative definite diagonal matrix. Hence, their sum, the Hessian matrix, is a negative definite matrix.
Therefore, we conclude that the canonical competition kernel will not cause any diversification and the population will remain monomorphic in its evolutionary journey.

That we get identical conditions for evolutionary branching in higher dimensions for the new class of stochastic processes, DMEP, as can be seen
by examining the components of the Hessian matrix, $H'_{ij}$ for both scenarios,

\begin{align}
 H'_{ij} = \bigg(\frac{\partial^2}{\partial y_i \partial y_j} f(\bold{x}, \bold{y})_{\big | \bold{y}=\bold{x}}\bigg).
\end{align}
which is equal to

\begin{widetext}
\begin{align}
\label{mod_hess}
H'(\bold{x}) = -m(m-1) \begin{pmatrix}  w_1 - u_1\\...  \\w_d - u_d\end{pmatrix} \begin{pmatrix}  w_1 - u_1, & ...,& w_d - u_d\end{pmatrix}
 + m H(\bold{x})
 \end{align}
\end{widetext}

which is clearly negative definite. 

As a natural extension of the canonical competition kernel, we modify it by adding a Gaussian component to it. The Gaussian competition kernel is biologically inspired since it is well known that individuals that are similar compete more strongly and this is captured by the Gaussian component.
Expressing the modified competition kernel as:
\begin{align}
 \label{MCC}
\alpha({\bold{x}, \bold{y}}) = \exp\Big(\sum_{i=1}^{d} w_i(\bold{x})(x_i - y_i)\Big) \exp\Big(\sum_i \frac{(x_i - y_i)^2}{2 \sigma_i^2}\Big),
\end{align}
the matrix elements of the Hessian of the invasion fitness function associated with it takes the form, 
\begin{widetext}
\begin{align}
 \label{GFEM}
\frac{\partial^2 f({\bold{x}, \bold{y}})}{\partial y_i \partial y_j} = - w_i(\bold{x}) w_j(\bold{x}) + w_i(\bold{x}) x_j^3 +w_j(\bold{x}) x_i^3 - 2 x_i^3x_j^3 -(3x_i^2-\frac{1}{\sigma_i^2})\delta_{ij},
\end{align}
\end{widetext}
and the Hessian matrix can be expressed as 
\\
\begin{widetext}
\begin{align}
 \label{GFHEM}
H(\bold{x}) = - \begin{pmatrix}  w_1(\bold{x}) - x_1^3 \\...\\w_d(\bold{x}) - x_d^3\end{pmatrix} \begin{pmatrix}  w_1(\bold{x}) - x_1^3 & ...& w_d(\bold{x}) - x_d^3\end{pmatrix} + \begin{pmatrix} -3x_1^2+\frac{1}{\sigma_i^2}&...&0\\...&...&...\\0&...&-3x_d^2+\frac{1}{\sigma_d^2}\end{pmatrix}.
\end{align}
\end{widetext}
Compared to the Hessian in Eq.\ref{FHEM} associated with the canonical competition kernel, the Hessian matrix above can possess positive eigenvalues depending on the variance of the Gaussian components, $\sigma_k$,
and the coordinates in the phenotypic space. Therefore, there are directions in the phenotypic space along which the coordinate $\bold{x}$ is a fitness minima and hence a potential evolutionary branching point. From the form of the Hessian matrix for the generalized logistics family of ecological models, and especially those corresponding to DMEP, as given by Eq. \ref{mod_hess}, we can see these models have a similar propensity of diversification upon adding a Gaussian competition kernel.

Hence we conclude that diversification under DMEP is expected to behave similar to that under SSIBM. Though we have derived and showed identical conditions for diversification for a monomorphic population, our simulations provide evidence that this in true even after the first branching point.

\section{Derivation and geometrical interpretation of the canonical equation from DMEP}
\label{deriv}
In this section we give a single step derivation of the canonical equation of adaptive dynamics based on DMEP from first principles. The derivation is a good starting point to appreciate 
the fundamental dynamics of evolution and we will be mainly interested in an intuitive and heuristic approach. Rigorous treatment of the standard adaptive dynamics limits like the infinite population sizes and infinitesimal mutations like  \cite{champagnat_etal2008} is not in the scope of this work. 
 Canonical equation can be viewed as dynamics that unravel as individuals go through the birth-death process described above. When a birth event is chosen, a new individual is added to the population with a phenotype that is offset from the parent by a small mutation chosen from a uniform distribution with a small amplitude. Death event results in the disappearance of an individual from the population. Therefore, geometrically speaking, one can view the birth/death events as the displacement of the centre of mass of the population cloud by a very small distance in the phenotypic space. The resulting dynamics represents the motion of a cloud of points in the phenotypic space. The points in the cloud represent individuals. The motion of the cloud is due to the appearance of new individuals with mutations due to birth events and the disappearance of individuals from the population cloud due to death events. We will first present arguments to interpret the canonical equation as it occurs in case of the logistic model, i.e. when the birth events are neutral. As we shall see it is easy to generalize this to the case when selection acts on both birth and death events.
 
Considering the logistic model, we consider a time period $ \Delta \tau$ which is very small compared to the total time in which the dynamics unravels but nonetheless large enough that we have a large number of birth and death events occurring. 
 The time, $\Delta \tau$, is the time in which a small displacement of the population cloud takes place due to birth/death events.
For large population sizes (and in the limit of infinite populations), we will have a large number of birth-death events happening even when $\Delta \tau$ becomes very small. We also assume the number of birth events $N_b$ will be equal to the number of death events $N_d$ in the time period $\Delta \tau$. This is because the population is assumed to be at equilibrium and therefore the probability of birth is equal to the probability of death event, i.e $P(Birth) = P(Death)$. Therefore, invoking the law of larger numbers, the sequence of birth-death events will be typical sequences with $N(birth) = N(death)$.
Consider a monomorphic population cloud at a particular point $\bold{x}$ in the phenotypic space. In the limit of very small mutations and a sufficiently small $ \Delta \tau$, a single birth event, $i$, will cause a displacement of the centre of mass of the population cloud given by $\Delta r_i$. Since for the logistic model, the birth events are neutral, and therefore for a large number of such events, $\sum_i{\Delta r_i} = 0$.  Similarly, the displacement by a single death event is, $\Delta q$, and the direction of displacement is in the opposite direction to the dying individual  $k$, whose death rate, by definition of DMEP, is given by  $ \delta_k =max(\delta_1, \delta_2, ..., \delta_n)$, or,
$\delta_{k}=max\big(\sum_{\j\neq \i}
\alpha(\bx_{i},\bx{_j})/K(\bx_{i})\big))$. Assuming all individuals except $k$ to be a monomorphic population, $\delta_{k}=max\big(
\alpha(\bx,\bx +\Delta q)/K(\bx +\Delta q)\big))$. The evaluation of the maximum is done in the following way, which also gives us the canonical equation. We seek to determine a vector displacement $\Delta q$ that will maximize $\delta_{k}$. By definition, the gradient of the function, $ \frac{ \alpha(\bold{x}, \bold{y}) K(\bold{x})}{K(\bold{y})}$, will be the direction that will maximize $\big(
\alpha(\bx,\bx +\Delta q)/K(\bx +\Delta q)\big))$.  That is, we pick a $k$ to die such that  $\Delta q$ is in the direction of the gradient of the function,  $ \frac{ \alpha(\bold{x}, \bold{y}) K(\bold{x})}{K(\bold{y})}$, with respect to $y$ at the value $x$.  We assume the same direction for $\Delta q$ for all the death events occurring in the time interval $\Delta \tau$.
Therefore, the total net displacement is given by
 \begin{align}
 \label{netdisplacement}
 \sum_i{\Delta r_i} + \sum_j{\Delta q_j} = N_{death}\nabla_{\bold{y} |\bold{y} = \bold{x}} f(\bold{x}, \bold{y}). |\Delta q|.
\end{align}
 Taking the limits, $N_{death} \rightarrow \infty$, $|\Delta q| \rightarrow 0$ and $\tau \rightarrow 0$
 we get the canonical equation of adaptive dynamics, 
 \begin{align}
 \label{AD_basic_der}
 \frac{d{\bold{x}}}{dt} = \gamma \nabla_{\bold{y} |\bold{y} = \bold{x}} f(\bold{x}, \bold{y}).
\end{align}

 In general and as mentioned above, birth rates might also depend on interactions between individuals and the fitness function will have a more general form involving competition kernels for both birth and death.  Let, $\alpha_{birth}({\bold{x}_j, \bold{x}_i)}$ and $\alpha_{death}({\bold{x}_j, \bold{x}_i)}$ be the competition kernels for birth and death respectively. Since we assume an infinite population limit at equilibrium we have
 $N(birth) = N(death)$. In this case, we will need to add up the displacement by both birth as well as death events. 
 For a single birth event, the displacement  is, $\Delta q_{b}$, and the direction of displacement is in the direction of the newly born individual  $k$, where  $ \rho_{k} =max(\rho_1, \rho_2, ..., \rho_n)$.
$\rho_{k}=max\big(\sum_{\j\neq \i}
\alpha_{birth}(\bx_{i},\bx{_j})/K(\bx_{i})\big))$. Assuming all individuals, except $k$, as a monomorphic population, $\rho_{k}=max\big(
\alpha_{birth}(\bx,\bx +\Delta q_b)/K(\bx +\Delta q_b)\big))$. Thus, we seek to determine a vector displacement $\Delta q_b$ that will maximize $\rho_{k}$. This is accomplished when $\Delta q_b$ is in the direction of the gradient of the function $ \frac{ \alpha_{birth}(\bold{x}, \bold{y}) K(\bold{x})}{K(\bold{y})}$ with respect to $y$ at the value $x$. We assume the same direction for $\Delta q_b$ for all the birth events occurring in the time interval $\Delta \tau$. The calculation for the net displacement, $\Delta q_d$, due to death events follows similarly though the displacement is in the opposite direction (and hence a negative sign) to the individual that dies. Therefore, the total net displacement, with $N(birth) = N(death)$, is given by
  
 \begin{align}
 \label{netdisplacement1}
 \sum_i{\Delta r_i} + \sum_j{\Delta q_j} = N_{death}\nabla_{\bold{y} |\bold{y} = \bold{x}} f'(\bold{x}, \bold{y}). |\Delta q|,
\end{align}
where, $ f'(\bold{x}, \bold{y}) = \Big( \frac{ \alpha_{birth}(\bold{x}, \bold{y}) K(x) }{K(\bold{y})} - \frac{ \alpha_{death}(\bold{x}, \bold{y}) K(x) }{K(\bold{y})}\Big)$ is the per capita growth rate of a mutant $y$ in a resident population. In other words, $ f'(\bold{x}, \bold{y})$ is the invasion fitness function and we recover canonical equation of adaptive dynamics.

We interpret the above equation as the translation of a population cloud along the direction of instantaneous gradient of the fitness function with respect to the mutant phenotype. It is remarkable that in this derivation, unlike the standard derivation \cite{dieckmann_law1996} which essentially considers the evolutionary trajectory as a sequence of invasions by the mutant phenotypes,
 we do not need to explicitly invoke concepts like ``invasion fitness" but recover it through the translation of the population cloud undergoing a birth-death process. 
 
Equation \ref{netdisplacement1} also suggests a faster way to implement birth death process when we have competition kernels for both birth and death events. At each iteration, we simply choose exactly one birth and one death event. For the birth event, the individual with the maximum birth rate is chosen and a new individual in its vicinity is produced. For the death event, as before, the individual with the maximum death rate is chosen to die. It is clear from the form of the invasion fitness function, Eq. \ref{fitness}, consisting of two separate terms for birth and death that such a stochastic model will satisfy the canonical equation of adaptive dynamics.
The advantage we get is faster simulations. Qualitatively speaking, evolution is a directed random walk in trait space determined by birth-death processes, selection and mutation. As the simulations in the next section reveal, DMEP causes higher competition, removes the ``jitters" in the trajectory and makes evolutionary trajectory more streamlined and at a faster pace than SSIBM

\section{Conclusion} 
Individual based models are essential in order to study various aspects of evolution like diversification,
predictability, adaptive radiation and frequency dependence. In this work, we derive the canonical equation of adaptive dynamics in a single step 
by introducing a new class of individual based models that coincide with the standard stochastic individual based models under maximal competition. We provide a very simple and intuitive way of understanding the correspondence between adaptive dynamics and individual based models.
 This can be regarded as a toy model to study how evolutionary trajectory unravels in the phenotypic space through microscopic birth-death processes. That we are able to derive the canonical equation of adaptive dynamics without explicitly invoking the master equation and concepts of invasion is of significance. 
 Previous work, including the seminal work by Dieckman and Law \cite{dieckmann_law1996}, links the fundamental microscopic ecological processes with adaptive dynamics using the concepts like invadability of a monomorphic population by nearby mutants with a higher invasion fitness. Experience with individual based simulations show a more realistic and fundamental picture where evolution is the motion of a population cloud in the phenotypic space whose center of mass is displaced due to selective birth and death events. This suggests, that the concept of fitness including invasion fitness is at most a derived quantity and not a fundamental ingredient of evolution. Our study 
 provides a mathematical structure to these ideas and provides a link between adaptive dynamics and individual based models by considering adaptive dynamics
 as the trajectory of the center of mass of the population cloud that unravels under maximal competition.

With the help of examples, we compare the rate of evolution under maximal and standard competition models .  We also demonstrate how DMEP can be interpreted as simulating the individual based model associated with a generalized logistic equation as given by Eq. \ref{modilo}, under appropriate limits, and how this yields exactly the same equations for adaptive dynamics as the standard logistic model. We also discuss how both processes yield exactly same condition for evolutionary branching in the deterministic limit. Our analysis gives an intuitive interpretation of the canonical equation of adaptive dynamics as the translation of the population cloud in the phenotypic space in the instantaneous direction of the gradient of the fitness function that arises due to birth-death events. A complete analysis of these connections is the focus of our future work.

Our simulations also enable us to address important questions regarding the nature of evolutionary dynamics.
  In \cite{doebeli_ispolatov2014}, it was shown that phenotypic properties can combine in complicated ways and as a result evolutionary trajectories can exhibit chaos. Evolution might be a complicated trajectory in the phenotypic space that is continuously twisting, turning and folding upon itself on a chaotic attractor. It is interesting to see that under maximal competition of DMEP, the adaptive dynamics shows similar biodiversity and can exhibit chaos, especially in high dimensional phenotypic spaces. Even more interesting, in our opinion, is the faster rate of evolution under maximal competition. This might be of importance from an ecological point of view, where such maximal competition might occur due to various abiotic factors like high temperature. We also saw that the adaptive dynamics based on the generalized family of logistic equations, Eq. \ref{modilo}, all give the same canonical equation upto a scaling factor. It is an interesting observation that ecological models given by the entire family of logistic like equations show very similar behavior as far as evolutionary dynamics is concerned. This interplay of ecology and evolution and ecology influencing the rate of evolution needs to be explored further.
  We hope our work contributes to the discussion on the concept of fitness in evolution and is useful for simulations involving birth death processes in high dimensional phenotypic spaces as well as the study of evolution of language, culture, religions and agent based models in economics.

\begin{acknowledgements}
VM acknowledges comments from Nicolas Champagnat and Michael Doebeli. VM greatly benefited from discussions with Michael Doebeli as well as his perspectives on evolution.
 VM acknowledges NFIG funds from IIT Madras for financial support. The author thanks Dr. Viswanadha Reddy for help with the videos.
\end{acknowledgements}




\bibliographystyle{plain}

\bibliography{EvolutionofDiversity2.bib}

\begin{thebibliography}{10}

\bibitem{champagnat_etal2006}
Nicolas Champagnat, R{\'{e}}gis Ferri{\`{e}}re, and Sylvie M{\'{e}}l{\'{e}}ard.
\newblock {Unifying evolutionary dynamics: From individual stochastic processes
  to macroscopic models}.
\newblock {\em Theoretical Population Biology}, 69(3):297--321, may 2006.

\bibitem{champagnat_etal2008}
Nicolas Champagnat, R{\'{e}}gis Ferri{\`{e}}re, and Sylvie M{\'{e}}l{\'{e}}ard.
\newblock {From Individual Stochastic Processes to Macroscopic Models in
  Adaptive Evolution}.
\newblock {\em Stochastic Models}, 24(sup1):2--44, nov 2008.

\bibitem{dieckmann_doebeli1999}
U.~Dieckmann and M.~Doebeli.
\newblock On the origin of species by sympatric speciation.
\newblock {\em Nature}, 400(6742):354--357, 1999.

\bibitem{dieckmann_doebeli2004}
U.~Dieckmann and M.~Doebeli.
\newblock Adaptive dynamics of speciation: Sexual populations.
\newblock In U.~Dieckmann, M.~Doebeli, J.~A.~J. Metz, and D.~Tautz, editors,
  {\em Adaptive Speciation}, Cambridge Studies in Adaptive Dynamics, pages
  76--111. Cambridge University Press, Cambridge, 2004.

\bibitem{dieckmann_law1996}
U.~Dieckmann and R.~Law.
\newblock The dynamical theory of coevolution: A derivation from stochastic
  ecological processes.
\newblock {\em J. Math. Biol.}, 34:579--612, 1996.

\bibitem{doebeli2011}
M.~Doebeli.
\newblock {\em Adaptive diversification}.
\newblock Princeton University Press, Princeton, 2011.

\bibitem{doebeli_ispolatov2014}
Michael Doebeli and Iaroslav Ispolatov.
\newblock {Chaos and unpredictability in evolution}.
\newblock {\em Evolution}, 68(5):1365--1373, mar 2014.

\bibitem{10.7554/eLife.23804}
Michael Doebeli, Yaroslav Ispolatov, and Burt Simon.
\newblock Point of view: Towards a mechanistic foundation of evolutionary
  theory.
\newblock {\em eLife}, 6:e23804, feb 2017.

\bibitem{geritz_etal1998}
S.~A.~H. Geritz, E.~Kisdi, G.~Mesz\'ena, and J.~A.~J. Metz.
\newblock Evolutionarily singular strategies and the adaptive growth and
  branching of the evolutionary tree.
\newblock {\em Evol. Ecol.}, 12(1):35--57, 1998.

\bibitem{gillespie1976}
D.~T. Gillespie.
\newblock General method for numerically simulating stochastic time evolution
  of coupled chemical-reactions.
\newblock {\em Journal of Computational Physics}, 22(4):403--434, 1976.

\bibitem{ISPOLATOV201697}
Iaroslav Ispolatov, Vaibhav Madhok, and Michael Doebeli.
\newblock Individual-based models for adaptive diversification in
  high-dimensional phenotype spaces.
\newblock {\em Journal of Theoretical Biology}, 390:97 -- 105, 2016.

\bibitem{ito_dieckmann2014}
Hiroshi~C Ito and Ulf Dieckmann.
\newblock Evolutionary branching under slow directional evolution.
\newblock {\em Journal of theoretical biology}, 360:290--314, 2014.

\bibitem{may1976}
R.~M. May.
\newblock Simple mathematical models with very complicated dynamics.
\newblock {\em Nature}, 261:459--467, 1976.

\bibitem{metz_etal1992}
J.~A.~J. Metz, R.~M. Nisbet, and S.~A.~H. Geritz.
\newblock How should we define fitness for general ecological scenarios.
\newblock {\em Trends in Ecology \& Evolution}, 7(6):198--202, 1992.

\bibitem{Note1}
The author acknowledges Michael Doebeli for this perspective.

\bibitem{Note2}
See Supplemental Material at [URL will be inserted by publisher] for examples
  and movies of simulations.

\end{thebibliography}

%
%

%
\end{document}